\documentclass[11pt]{article}
\pdfoutput=1
\usepackage{jcapmod}
\usepackage{booktabs}
\usepackage[english]{babel}
\usepackage{amsmath,amssymb,amsbsy,amstext, amsthm, simplewick}
\usepackage{graphicx}
\usepackage{amsfonts}
\usepackage{amssymb}

\usepackage{upgreek}
 \usepackage{exscale,relsize}
\allowdisplaybreaks[1]

\newcommand{\vrad}{v_{\text{r}}}
\newcommand{\dm}{\rho_{\text{DM}}}

\newcommand{\rad}{\rho_{\text{r}}}
\newcommand{\M}{\text{M}_{\text{pl}}}
\newcommand{\G}{\text{G}}
\newcommand{\vp}{\varphi}
\newcommand{\be}{\begin{equation}}
\newcommand{\ee}{\end{equation}}
\newcommand{\bea}{\begin{eqnarray}}
\newcommand{\eea}{\end{eqnarray}}

\def\lp{\left(}
\def\rp{\right)}
\def\lb{\left[}
\def\rb{\right]}

\usepackage{amsmath}
\usepackage{amssymb}
\usepackage{epsfig}
\usepackage{graphicx}
\usepackage{multirow}
\usepackage{color}
\usepackage[normal]{subfigure}
\usepackage{rotating}
\usepackage{hyperref}

\begin{document}

\begin{titlepage}

\setcounter{page}{1} \baselineskip=15.5pt \thispagestyle{empty}

\bigskip\

\vspace{1cm}
\begin{center}
{\fontsize{19}{28}\selectfont  \sffamily \bfseries Modified Dust and the Small Scale Crisis in CDM}

\end{center}

\vspace{0.2cm}

\begin{center}
{\fontsize{13}{30}\selectfont  Fabio Capela$^{a}$ and Sabir Ramazanov$^{b}$}
\end{center}

\begin{center}

\vskip 8pt
\textsl{$^a$ DAMPT, Centre for Mathematical Sciences, Cambridge University,\\ 
Wilberforce Road, Cambridge CB3 0WA, UK}\\
\textsl{$^b$ Service de Physique Th\'{e}orique, Universit\'{e} Libre de Bruxelles (ULB), \\
CP225 Boulevard du Triomphe, B-1050 Bruxelles, Belgium}
\vskip 7pt

\end{center}

\vspace{1.2cm}
\hrule \vspace{0.3cm}
\noindent {\sffamily \bfseries Abstract} \\[0.1cm]
At large scales and for sufficiently early times, dark matter is described 
as a pressureless perfect fluid---dust---non-interacting with Standard Model fields. 
These features are captured by a simple model with two scalars: a Lagrange multiplier and 
another playing the role of the velocity potential. 
That model arises naturally in some gravitational frameworks, e.g., the mimetic 
dark matter scenario. We consider an extension 
of the model by means of higher derivative terms,
such that the dust solutions are preserved 
at the background level, but there is a non-zero sound speed at the linear level. 
We associate this {\it Modified Dust} with 
dark matter, and study the linear evolution of cosmological perturbations in that picture. 
The most prominent effect is the suppression of their power 
spectrum for sufficiently large cosmological momenta. This 
can be relevant in view of the problems that cold dark matter faces at sub-galactic scales, 
e.g., the missing satellites problem. At even shorter scales, 
however, perturbations of Modified Dust are enhanced compared to the predictions 
of more common particle dark matter scenarios. This is a peculiarity of their evolution 
in radiation dominated background. We also briefly discuss clustering 
of Modified Dust. We write the system of equations in the Newtonian 
limit, and sketch the possible mechanism which could prevent 
the appearance of caustic singularities. The same mechanism may be 
relevant in light of the core-cusp problem. 
\vspace{0.2cm}
\hrule
\hrule

\vspace{0.6cm}
 \end{titlepage}

\hrule
 \tableofcontents
\vspace{0.2cm}
\hrule
\hrule

\vspace{0.6cm}

\section{Introduction}
 
To date, particle physics provides us with a plethora of candidates for dark matter~(DM). 
These include sterile neutrinos, supersymmetric partners of the Standard Model particles, 
light scalars (axions) and many others.  
Given that the new degrees of freedom are heavy enough and weakly interacting with the 
Standard Model constituents, one deals with cold dark matter (CDM). The concept of CDM is 
of the uttermost importance in modern cosmology. 
Namely, it is among the building blocks of the 6-parametric concordance model established 
by the recent Planck and WMAP missions~\cite{2013ApJS..208...19H,2013arXiv1303.5076P}. 
In particular, the evolution of linear perturbations developed in this framework 
is in excellent agreement with the picture of the Cosmic Microwave Background 
temperature anisotropies~\cite{Ade:2013ktc}. Furthermore, simulations based on CDM 
lead to correct predictions about the large scale structure distribution 
in the Universe~\cite{2001MNRAS.327.1297P,Eisenstein:2005su,2004ApJ...606..702T}. 
Finally, the concept of CDM agrees with the 
Bullet Cluster observations~\cite{2006ApJ...652..937B,Clowe:2006eq} 
and provides an explanation for 
the flat galaxy rotation curves~\cite{Navarro:1995iw,Navarro:1996gj,2009arXiv0901.0632E}.

Despite these successes, there is some tension between CDM predictions 
and astronomical data at the sub-galactic scales. This amounts to three 
problems. First, CDM predicts an overabundance of small structures, i.e., dwarf galaxies. 
Observations in the vicinity of the Milky Way, 
however, indicate a much smaller number. 
This states the ``missing satellites problem''~\cite{Klypin:1999uc,Moore:1999nt}. Second, CDM 
leads to cuspy profiles of the DM halos~\cite{Navarro:1995iw, Navarro:1996gj}. At the same 
time, observations on the concentrations of dwarf galaxies rather prefer cored profiles~\cite{McGaugh:2002vi,2009MNRAS.397.1169D}. 
Finally, N-body simulations result into a large central density of 
massive subhaloes in the Milky Way---a fact, which 
is in conflict with observations of stellar dynamics in dwarf galaxies hosted by those subhaloes.
This inconsistency is dubbed as the ``too big to fail'' problem~\cite{2011MNRAS.415L..40B}. It is 
important to keep in mind that the aforementioned shortcomings of CDM may not be robust 
to the proper account of various astrophysical 
phenomena~\cite{Benson:2001au, Arraki:2012bu,2013ApJ...765...22B,Somerville:2001km,Bullock:2000wn,
Dekel:2002yh,Governato:2006cq,Mashchenko:2006dm,Pontzen:2011ty}. 
However, realistic high resolution numeric simulations, 
which include baryonic processes, are challenging to implement at the moment. 
Meanwhile, it is interesting to speculate if part or all of the problems are due 
to the peculiar nature of DM itself. For example, Warm Dark Matter (WDM) scenarios 
start from the proposal that the DM particles are still mildly relativistic at the freeze out 
temperature. Then, the short wavelength perturbations get washed out due to free streaming 
processes~\cite{Bode:2000gq,Colin:2000dn}. This provides a simple mechanism to 
suppress the number of small scale structures, and thus can be relevant to alleviate the missing satellites 
problem~\cite{Goetz:2002vm}. However, WDM scenarios are perhaps too efficient in erasing 
the short wavelength perturbations, as indicated by the constraints following from the 
Lyman-$\alpha$ forest data~\cite{Boyarsky:2008xj}. Another approach 
to the problems at the sub-galactic scales is to go beyond the 
approximation of collisionless matter. Namely, allowing for sufficiently strong self-interactions of 
DM particles~\cite{Spergel:1999mh}, one can address the core-cusp and too big to fail problems~\cite{2014arXiv1412.1477E}. 
On the other hand, constraints obtained from the Bullet Cluster observations imply 
smaller self-interaction cross-sections than what is required in view of the core-cusp problem~\cite{Randall:2007ph}.

One can try to find a solution to the small scale crisis by 
switching to a paradigm different from particle DM. 
Recently, an interesting proposal on the way to model dark matter and dark energy 
has been made in Ref.~\cite{Lim:2010yk}. There, the authors introduced a novel class of theories, 
referred to as the $\Sigma \varphi$-fluid. 
The action for the $\Sigma \varphi$-fluid is given by
\begin{equation}
\label{actiongen}
\mathcal{S}= \int d^4x \sqrt{-g} \left[\Sigma \left(g^{\mu \nu}
\partial_{\mu} \varphi \partial_{\nu}\varphi  -1\right) +K(\varphi, \partial_{\mu} \varphi) \right] \; .
\end{equation}
In what follows, we assume the sign convention $(+,-,-,-)$ for the metric. 
Here $\Sigma$ is the Lagrange multiplier; 
$K(\varphi, \partial_{\mu} \varphi)$ is some arbitrary function of the scalar $\varphi$ 
and its derivatives. 
Varying the action~\eqref{actiongen} with respect to the field $\Sigma$ enforces the 
constraint $g^{\mu \nu} \partial_{\mu} \varphi \partial_{\nu} \varphi=1$, so that 
$\partial_{\mu} \varphi$ is a unit 4-vector. One can associate the field $\varphi$ 
with the velocity potential of the $\Sigma \varphi$-fluid. The constraint equation 
tells us that the fluid elements follow geodesics, much 
in the same manner as the dust particles. At the same time, given the non-trivial 
function $K(\varphi, \partial_{\mu} \varphi)$, one can allow for 
a non-zero effective pressure with a time-dependent equation of state. This opens 
up the possibility to construct a fluid, which may mimic dust at early 
times and a positive cosmological constant $\Lambda$ later on. 

In the present paper, however, we restrict the discussion 
to DM. A pressureless perfect fluid is obtained for the function $K(\varphi, \partial_{\mu} \varphi)$ identically equal to zero, i.e.,
 $K(\varphi, \partial_{\mu} \varphi)=0$. The energy density of the dust is associated with the field $\Sigma$ 
decaying as $1/a^3$ with the scale factor $a$. It is thus 
tempting to view the 
construction with the Lagrange multiplier as the simplest model of DM: at the background and linear levels, it reproduces 
all the successes of more common particle scenarios. In the non-linear regime, however, the dust model has important drawbacks: it leads to 
caustic singularities and is unable to form stable DM halos~\cite{Sahni:1995rm}. 

From now on, we should go beyond the approximation of 
a pressureless perfect fluid. In the theory described by the action~\eqref{actiongen}, this is achieved by incorporating 
the function $K (\varphi, \partial_{\mu} \varphi)$. At the same time, we would like to keep 
the simple form of the background dust solutions in the Universe 
dominated by the $\Sigma \varphi$-fluid: $\Sigma \propto 1/a^3$. 
Interestingly, this is possible with the non-trivial choice of the function 
$K (\varphi, \partial_{\mu} \varphi)$,
\begin{equation}
\label{K}
K(\varphi, \partial_{\mu} \varphi) \equiv K(\partial_{\mu} \varphi)=\frac{\gamma_1}{2} 
\nabla_{\mu} \nabla^{\mu} \varphi \nabla_{\nu} \nabla^{\nu} \varphi 
+ \frac{\gamma_2}{2} \nabla_{\mu} \nabla_{\nu} \varphi \nabla^{\mu} \nabla^{\nu} \varphi \; .
\end{equation}
Here $\gamma_1$ and $\gamma_2$ are parameters with dimension of mass squared. 
Expressions~\eqref{actiongen} 
and~\eqref{K} state the model, which we refer to as Modified Dust in what follows. We 
shortly give a brief summary of our main results. 
Before that, let us comment on how the action~\eqref{actiongen} 
arises in different gravitational frameworks.

In Ref.~\cite{Chamseddine:2013kea}, the authors considered the standard 
Einstein's metric  $g_{\mu \nu}$ as the composition of an 
auxilliary metric $\tilde{g}_{\mu \nu}$ and the first derivatives of the scalar field $\vp$, 
\begin{equation}
\nonumber 
g_{\mu \nu} =\tilde{g}_{\mu \nu} \tilde{g}^{\alpha \beta} \partial_{\alpha} \vp \partial_{\beta} \vp .
\end{equation}
Unexpectedly, variation with respect to the fields $\tilde{g}_{\mu \nu}$ and $\vp$
results in a modification of the Einstein--Hilbert equations such that the traceless part is 
non-zero even in the absence of matter. 
This discrepancy with the standard equations of General Relativity is due to the presence 
of an extra degree of freedom, which behaves as dust. The dust solution has been dubbed 
as mimetic dark matter in Ref.~\cite{Chamseddine:2013kea}.
Soon afterwards, it has been realized that the proposed scenario is 
equivalent to adding a Lagrange multiplier 
into the Einstein--Hilbert action~\cite{Golovnev:2013jxa}. This is the picture 
described by the action~\eqref{actiongen}. 
Furthermore, in Ref.~\cite{Barvinsky:2013mea} it was proved that the 
condition $\Sigma >0$ set on the initial Cauchy surface is sufficient to avoid ghosts. 
The concept of mimetic dark matter could be 
interesting from at least two perspectives. First, it is in direct 
relation with the broken disformal transformations in gravity~\cite{Deruelle:2014eha}. 
Quite surprisingly, mimetic dark matter also arises 
in the context of non-commutative geometry, which might be a promising setup for quantum 
gravity~\cite{Chamseddine:2014nxa}. Strictly speaking, the 
function $K(\varphi, \partial_{\mu} \varphi)$ 
equals to zero in the mimetic dark matter scenario as it stands. 
Setting it by hands, however, opens up the possibility to mimic different types of 
cosmologies~\cite{Chamseddine:2014vna,Saadi:2014jfa}.

Constructions introducing 
the term with the Lagrange multiplier are known in the context of 
Einstein--\AE ther (EA) models~\cite{Jacobson:2000xp}. In particular, scalar 
EA~\cite{Haghani:2014ita} has exactly the form given by Eqs.~\eqref{actiongen} and~\eqref{K}. 
This is interesting, as the scalar EA appears in the IR limit 
of the projectable version of Horava--Lifshitz 
model~\cite{Blas:2009qj,Jacobson:2014mda,Blas:2009yd}---
power counting renormalizable theory of gravity~\cite{Horava:2009uw}. It is thus 
not a surprise that DM has been identified in this setup~\cite{Mukohyama:2009mz}. 
Contrary to the case of Modified Dust, however, it was suggested in~\cite{Mukohyama:2009mz,Mukohyama:2009tp} 
to extend the pressureless perfect fluid by means of higher curvature 
terms inherent to Horava's proposal.  

In the present paper we prefer to stay on a more phenomelogical side. 
Our main purpose is to study the linear evolution of cosmological perturbations 
of Modified Dust described by Eqs.~\eqref{actiongen} and~\eqref{K}. 
We uncover several effects, which can be relevant in light 
of the small scale crisis. The first is the suppression of perturbations with relatively large momenta. 
As we will see explicitly in Section~\ref{sec:einstein-de-sitter}, $\gamma$-terms result 
into a non-zero sound speed at the level 
of perturbations~\cite{Chamseddine:2014vna}. Beyond the sound horizon, perturbations of Modified Dust 
behave as in the CDM picture: they grow linearly with the scale factor during 
the matter dominated (MD) stage. 
As they enter the sound horizon, their growth stabilizes. 
This places an important cutoff on the linear power spectrum in our model. In this regard, 
Modified Dust carries similarities with WDM. So, by setting 
$\gamma_i \sim 10^{-10} \M^2$,  where $\M$ is the Planck mass\footnote{Hereafter, we define the Planck mass squared as 
the inverse of the Newton's constant $G$, i.e., $\M^2 =G^{-1}$.}, one can suppress perturbations with wavelengths below $100~\mbox{kpc}$. Hence, Modified Dust is capable to address the 
missing satellites problem. There is, however, an important distinction from the 
case of WDM scenarios. The difference is clearly seen from 
the evolution during the radiation dominated (RD) stage, 
when perturbations of Modified Dust 
of {\it very} small wavelengths experience a linear growth with the scale factor. 
We show this explicitly in Section~\ref{sec:inclusion-radiation}. As a result, corresponding 
perturbations get amplified compared to the predictions of WDM scenarios (and even CDM scenarios for sufficiently large redshifts). The effect, however, 
is only prominent for wavelengths in the pc-range. Therefore, its possible physical applications remain unclear at the moment.

Finally, in Section~\ref{sec:non-linear-level} 
we discuss the behaviour of Modified Dust in the non-linear 
regime. This is particularly relevant in light of caustic singularities 
occuring in the case of pressureless perfect fluid.
We notice, however, that deviations of Modified Dust from its 
conventional counterpart become prominent exactly where one expects the 
appearence of singularities. Based on this simple observation, we 
propose a mechanism that could be relevant to reduce 
the energy density in the dangerous regions. 

The outline of the paper is as follows. In Section~\ref{sec:einstein-de-sitter}, 
we discuss the evolution of the Universe 
filled in with Modified Dust. This we do at the background level and at the level of 
linear perturbations. In Section~\ref{sec:inclusion-radiation}, we discuss peculiarities 
of the linear evolution at very 
early times, during the RD stage. 
In Section~\ref{sec:non-linear-level}, we get back 
to the MD Universe and write down the relevant system of cosmological equations 
in the Newtonian limit. There, we discuss the clustering issues of Modified Dust. We finish 
by formulating some opened issues in the last Section.

\section{Matter Dominated Universe}
\label{sec:einstein-de-sitter}
We start with the case of the Universe dominated by Modified Dust. 
This is a good approximation to the Universe at relatively low redshifts, 
i.e., $z \ll 10^{4}$, when the contribution 
of the radiation can be neglected. Non-trivial effects arising upon the inclusion 
of radiation will be considered in the next Section. In the present paper, 
we will always neglect the contributions from baryons and 
dark energy. The former is expected to change the behaviour of the 
gravitational potential at the percent level, while the latter 
becomes relevant only at very small redshifts, $z \lesssim 1$. 

\subsection{Case $K(\partial_{\mu} \varphi)=0$}

In the simplest case, when the constants $\gamma$ equal to zero, 
i.e., $\gamma_1=\gamma_2=0$, we have for the 
energy momentum tensor of the $\Sigma \varphi$-fluid, 
\begin{equation}
\nonumber 
T^{\mu}_{\nu} =2\Sigma \partial^{\mu} \varphi \partial_{\nu} \varphi\; .
\end{equation}
Varying the action~\eqref{actiongen} with respect to the 
field $\Sigma$ (Lagrange multiplier), one obtains
\begin{equation}
\label{constraint}
\partial_{\mu} \varphi \partial^{\mu} \varphi =1 \; .
\end{equation}
We see explicitly that the energy momentum 
tensor is the one of a pressureless perfect fluid, 
with $2\Sigma$ being the energy-density and $\varphi$ 
the velocity potential. At the background level, the former drops with the scale factor $a$ 
as $2\Sigma = C/a^3$. We fix the constant $C$ in a way that $2\Sigma$ corresponds to the 
energy density of DM, i.e., 
\begin{equation}
\label{convention}
2 \Sigma =\rho_{DM} \; .
\end{equation}
The background value of the field $\varphi$ is given by 
\begin{equation}
\nonumber 
\varphi=t \; ,
\end{equation}
up to an irrelevant constant of integration. The cosmological 
evolution of the Universe filled in with dust is then obtained 
from the $(ij)$-component of Einstein's equations, 
\begin{equation}
\label{ijstand}
2{\cal H}' +{\cal H}^2=0 \; .
\end{equation}
Here ${\cal H}=a (\eta) H$, where $H$ denotes the Hubble parameter. 
From this point on we prefer to work in 
terms of the conformal time $\eta$. The prime denotes the 
derivative with respect to the latter. From Eq.~\eqref{ijstand} we obtain 
${\cal H} =2/\eta$, which gives the standard solution for the scale factor $a(\eta)$ 
in the matter dominated Universe: 
$a(\eta) \propto \eta^2$.  

Note one generic feature of the model, where the 
4-velocity of the fluid $u_{\mu}$ is the derivative of the scalar field $\vp$, i.e., $u_{\mu} =\partial_{\mu} \varphi$. 
In that case, the conservation of the energy-momentum tensor 
implies just one equation. Indeed, 
\begin{equation}
\nonumber
\nabla_{\mu} T^{\mu}_{\nu} =\nabla_{\mu} (2\Sigma \partial^{\mu} \varphi) 
\partial_{\nu} \varphi =0 \; ,
\end{equation}
where we took into account Eq.~\eqref{constraint}. 
Consistently, this equation can be obtained from the variation of the action~\eqref{actiongen} 
with respect to the field $\varphi$. Of course, this does 
not lead to degeneracy of solutions, as the constraint~\eqref{constraint} itself 
plays the role of the missing equation~\cite{Chamseddine:2014vna}. 

At the linear level, Eq.~\eqref{constraint} reads 
\begin{equation}
\label{euler}
\delta \varphi' =a \Phi \; .
\end{equation}
This is essentially the Euler equation linearized, 
where $\Phi$ is the scalar perturbation of the $(00)$-component of the 
metric. Hereafter, we choose to work in the Newtonian gauge. As it follows 
from the form of the action~\eqref{actiongen}, Eq.~\eqref{euler} remains unmodified even 
for the non-trivial choice of the function $K (\partial_{\mu} \varphi)$. 
In particular, the same equation is true for the case of non-zero 
coefficients $\gamma_1$ and $\gamma_2$, to which we will turn soon.

Before that, let us remind the picture of linear evolution in the 
case of the pure dust (CDM). In the Newtonian gauge, 
the perturbed Friedmann--Robertson--Walker metric has the form,
\begin{equation}
\nonumber 
ds^2 =a^2 (\eta) [(1+2\Phi)d\eta^2 -(1-2\Psi) d{\bf x}^2] \; .
\end{equation}
In the perfect fluid approximation, the simple relation holds 
between the potentials $\Phi$ and $\Psi$: $\Phi=\Psi$. 
We choose to work with the function $\Phi$ in what follows. 
The evolution of the potential $\Phi$ can be easily inferred 
from the $(ij)$-component of Einstein's equations,
\begin{equation}
\nonumber 
\Phi''+\frac{6}{\eta} \Phi'=0 \; .
\end{equation}
Up to a negligible decaying mode, it has the 
constant solution $\Phi=\text{const}$. 
In the presence of the constant gravitational potential, perturbations 
of DM grow linearly with the scale factor, as it immediately follows from the 
Poisson equation, 
\begin{equation}
\label{Poissonlin}
-k^2 \Phi =4\pi G a^2\dm \delta_{\text{DM}} \; ,
\end{equation} 
 where $G$ is the Newton's constant. That is, during the MD stage, perturbations of 
DM are subject to the Jeans instability independently of their momenta. 
Later on, they enter non-linear regime, and the clustering starts. This standard picture 
changes drastically upon the inclusion of $\gamma$-terms.

\subsection{Generic case}

Let us make two important comments before we dig into the details of calculations. 
With no loss of generality, we choose to work with the unique 
coefficient $\gamma_1$ in the bulk of the paper, while setting 
the other one, $\gamma_2$, to zero. Besides considerations of simplicity, 
this is also justified, since both terms are expected to result into qualitatively the same phenomenology. 
We put the details regarding the case $\gamma_2 \neq 0$ in the Appendix~\ref{sec:term2}. With this 
said, the energy momentum tensor has the form~\cite{Chamseddine:2014vna}, 
\begin{equation}
T^{\mu}_{\nu} =2\Sigma \partial_{\nu} \varphi \partial^{\mu} \varphi 
+\gamma \left(\partial_{\alpha} \varphi \partial^{\alpha} \square \varphi +
\frac{1}{2} (\square \varphi)^2 \right) \delta^{\mu}_{\nu}
 -\gamma \left(\partial_{\nu} \varphi \partial^{\mu} \square \varphi 
+\partial_{\nu} \square \varphi \partial^{\mu} \varphi \right) \; ,
\end{equation}
where $\gamma \equiv \gamma_1$. 

For non-zero coefficients $\gamma_1$ and $\gamma_2$ the interpretation of the 
field $\Sigma$ as the energy-density of DM may be misleading. Still, 
we prefer to stick to the simple convention~\eqref{convention} in what follows. 
Hopefully, this is not going to confuse the reader, 
as one is interested in the behaviour of the gravitational potential in the end. 
Moreover, the difference between the 
formal energy density and the physical one, $\dm^{\text{ph}}=T^{0}_{0}$, remains small in 
most cases discussed in the present paper. This is true at both levels of background and linear perturbations. 
The only exceptional case occurs at very early times, deeply in radiation dominated era. We will 
comment on that in due time. 

Let us write the background cosmological equations. The simplest is the 
one corresponding to the $(ij)$-component of Einstein equations,  
\begin{equation}
\left( 2{\cal H}'+{\cal H}^2 \right) \cdot \left( 1-12\pi G \gamma \right)=0\; .
\end{equation}
Apart from the degenerate case $12\pi G \gamma=1$, we have ${\cal H}=2/\eta$, which corresponds 
to the Universe driven by dust. Consistently, the conservation equation has the form,  
\begin{equation}
\label{consback}
\dm'+3{\cal H}\dm =\frac{3}{2}\gamma \left(2\frac{{\cal H}'}{a^2} 
+\frac{{\cal H}^2}{a^2} \right)' \; .
\end{equation}
The Friedmann equation is given by
\begin{equation}
\nonumber 
3 {\cal H}^2\left( 1-24\pi \gamma \G \right) =8\pi G a^2 \dm 
\end{equation}
Upon the formal change $(1-24\pi \gamma G)^{-1}\dm \rightarrow \dm$, 
we get back to the standard set of equations of the dust dominated Universe. 
In Appendix~\ref{sec:term2}, we show that the same conclusion holds 
for the $\gamma_2$-term in Eq.~\eqref{K}.

Before turning to the study of linear perturbations, let us make one useful observation. 
We note that the linearized energy-momentum tensor is similar to that of a perfect fluid 
in a sense that $\delta T^{i}_{j} \propto \delta^{i}_{j}$. 
This means that one can impose the constraint on the 
scalar perturbation $\Psi$ of the spatial part of the metric, $\Psi=\Phi$
\footnote{The analogous conclusion is 
not applicable to the $\gamma_2$-term. However, the difference 
between potentials $\Phi$ and $\Psi$ is negligible and hence is irrelevant for phenomenology. See details in Appendix~\ref{sec:term2}. We thank A.~Vikman 
for discussions on this point.}. We choose to work with the potential $\Phi$ in what follows.

\begin{figure}[t!]
\begin{center}
\includegraphics[width=0.70\columnwidth]{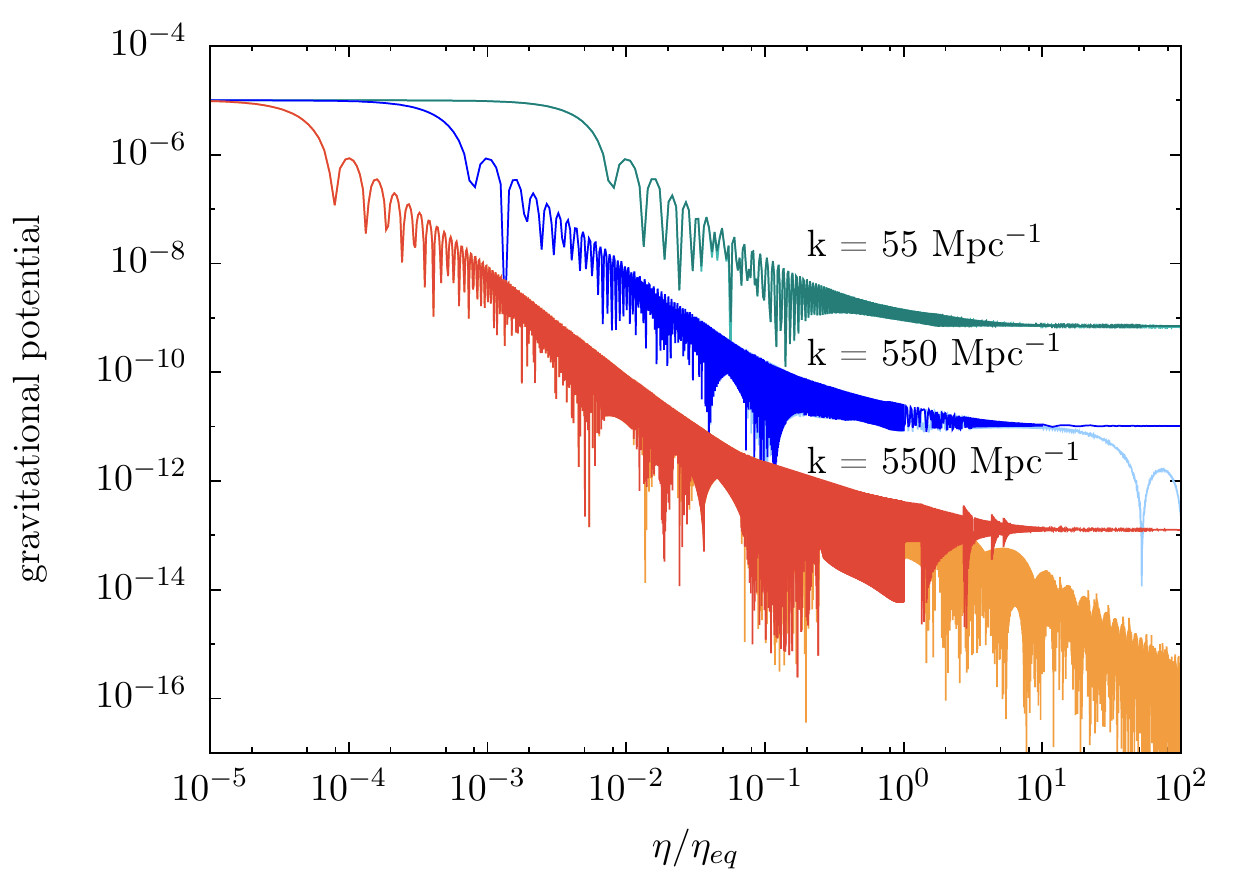}
\end{center}
\caption{ The gravitational potential $\Phi$ is plotted as a function of conformal time $\eta$ for 
different cosmological 
wavenumbers $k$. Cases of Modified Dust and a pressureless perfect fluid/ CDM have been studied. The 
time $\eta_{eq}$ corresponds to the equilibrium between matter and radiation. 
For the wavenumber $k=55~\mbox{Mpc}^{-1}$ (green line) predictions of Modified Dust and CDM 
are indistinguishable from each other. Shorter wavelength perturbations 
are pictured with dark blue and red lines (CDM) and light blue and 
orange lines (Modified Dust). The parameter $\gamma$ is set to $\gamma =10^{-10}~\M^2$.
\label{potential}}
\end{figure}

The simplest way to proceed is to write the $(0i)$-component 
of Einstein's equations. The reader can find the explicit 
expression in the Appendix~\ref{sec:term-gamma_1}. Here we write it in the approximation of the small 
parameter $\gamma$, i.e., $\gamma \ll \M^2$, 
\begin{equation}
\label{0iapprox} 
\delta \varphi''  +\left(c^2_s k^2 -\frac{3}{2} {\cal H}^2 \right) \delta \varphi =0 \; ,
\end{equation}
where $c^2_s$ is given by
\begin{equation}
\nonumber 
c^2_s \approx 4\pi \gamma \G \; .
\end{equation}
Clearly, the second term in the equation mimics the sound 
speed~\footnote{This observation was first made in the 
context of inflation~\cite{Chamseddine:2014vna}.}. This explains the notation ``$c^2_s$'' 
we use. Provided that the cosmological modes are beyond the speed horizon, i.e., 
$c^2_s k^2 \ll {\cal H}^2$, one obtains for the field 
$\varphi$ perturbation $\delta \varphi \propto \eta^3$. 
Using Eq.~\eqref{euler}, we get $\Phi =\mbox{const}$. 
This is the standard solution for the gravitational potential in the MD Universe. 
Another story occurs after the modes enter the sound horizon, i.e., 
in the regime $c^2_s k^2 \gg {\cal H}^2$. Accordingly to 
Eq.~\eqref{0iapprox}, the rapid growth of the perturbations of the 
field $\varphi$ stops and turns into oscillations, i.e., 
$\delta \varphi \propto e^{i c_s k \eta}$. As a result, the gravitational potential 
decreases with the scale factor.

The behaviour of the DM energy density perturbations can be easily deduced from the 
$(00)$-component of Einstein's equations, which takes the standard 
form~\eqref{Poissonlin} in the small $\gamma$ approximation. 
As it follows, for modes with sufficiently 
short wavelengths and at relatively late times, the linear 
growth of the energy density contrast $\delta_{\text{DM}}$ stabilizes and turns 
into oscillations with a constant amplitude. Let us choose the 
constant $\gamma$ in such a way that the growth stops at redshifts as small as $z \simeq 10$ 
for perturbations with the comoving wavelengths of $\lambda \simeq 100~\mbox{kpc}$. 
These wavelengths roughly characterize the collection regions collapsing to 
the halos of dwarf galaxies. 
The redshifts $z \simeq 10$ correspond to the times, when the modes 
of interest enter the non-linear regime (in the CDM picture). 
The estimate of the parameter $\gamma$ reads $\gamma \sim 10^{-10} \M^2$, 
which is remarkably close to the Grand Unification Scale. 
This value of the parameter $\gamma$ is required to alleviate the 
missing satellites problem. For larger values, one risks to 
affect the evolution of the galaxies in an unaffordable manner. 
Much smaller values $\gamma \ll 10^{-10} \M^2$ are, however, 
plausible, as the proper account of baryonic processes may eliminate the problem 
with the dwarf galaxies. In that case, however, the 
motivation for Modified Dust is essentially lost. Therefore, 
we set $\gamma \sim 10^{-10} \M^2$ in what follows, 
unless the opposite is stated. 

\begin{figure}[t!]
\begin{center}
\includegraphics[width=0.70\columnwidth]{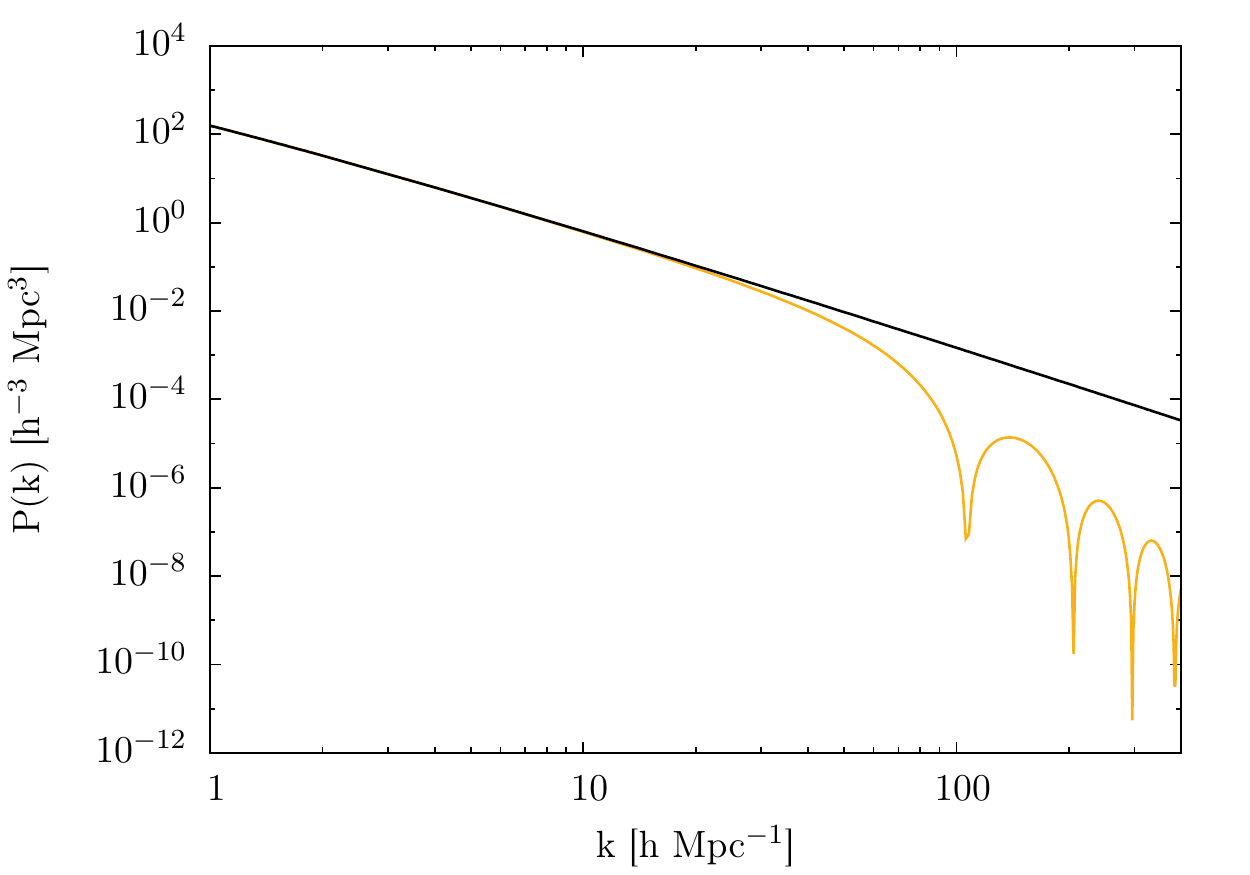}
\end{center}
\caption{The linear matter power spectrum as a function of the momenta $k$. Black and orange lines correspond 
to the cases of a pressureless perfect fluid/CDM and Modified Dust, respectively. The choice of the 
parameter $\gamma=10^{-10}~\M^2$ is assumed.
\label{power}}
\end{figure}

To put things on solid ground, we performed numerical simulations to obtain the 
gravitational potential and the 
matter power spectrum. The results are presented in Figs.~\ref{potential} 
and~\ref{power}, respectively. 
As it is clearly seen from Fig.~\ref{potential}, the gravitational potential 
deviates from the standard behaviour predicted 
by the CDM scenarios for sufficiently large cosmological momenta. 
Namely, at some point in MD stage it starts to oscillate with a decreasing amplitude 
and a frequency $\omega =c_s k$. The corresponding period of oscillations is 
very large: for wavelengths $\lambda \sim 100~\mbox{kpc}$ it is 
comparable with the age of the Universe. 
The decrease of the gravitational potential with the scale factor 
translates into the suppression of the linear matter power spectrum, 
which we plot in Fig.~\ref{power} for the redshift value $z=3$. 
The choice of the redshift is dictated by simplicity considerations: 
for smaller values of $z$, we would need to incorporate 
the effects of the accelerated expansion of the Universe 
into the analysis. Note that the plot in Fig.~\ref{power} represents the extrapolation of the linear 
evolution of Modified Dust to late times. This is by no means justified, as the cosmological modes of interest are 
deeply in the non-linear regime at the redshift $z=3$. Still, the plot is useful for the purpose of comparison 
with the particle DM scenarios. 
We use the conventional definition of the power spectrum, 
\begin{equation}
\nonumber 
P(k)=2\pi^2 \frac{\Delta^2 (k)}{k^3} \; .
\end{equation}
Here $\Delta (k)$ is the amplitude of the matter perturbations related to the 
two-point correlator in the coordinate space by 
\begin{equation}
\nonumber 
\langle \delta^2 ({\bf x})  \rangle  =\int \frac{dk}{k} \Delta^2 (k) \; .
\end{equation}
The picture~\ref{power} is quite analogous to what one has in the case 
of WDM scenarios~\cite{2012MNRAS.421...50V,Gorbunov:2008ui}. 
There are two important qualifications, however. While WDM models predict the exponential 
suppression of the small scale spectrum, we observe 
a more moderate power law drop. This distinction might be relevant, 
since WDM does perhaps too good 
job with diluting sub-galactic structures. Second, we note the presence of slow oscillations in 
the high momentum tail of the Modified Dust spectrum. 
This could be a promising smoking gun of our scenario. 

The conclusions of the present Section work well for wavelengths 
in the $\mbox{kpc}$-range. For perturbations with even smaller wavelengths another effect 
becomes prominent. That is, very small perturbations get enhanced 
during the evolution in the RD epoch (once again, compared to the 
predictions of CDM). We discuss this issue in the following Section.

\section{Inclusion of radiation}
\label{sec:inclusion-radiation}
\subsection{Initial conditions}

In the presence of radiation, the evolution of Modified Dust changes considerably. In particular, 
the energy density associated with the field $\Sigma$ is now given by 
\begin{equation}
\label{dust-rad} 
\dm =\rho_{0} +\tilde{\rho}_{\text{r}} \; , 
\quad \tilde{\rho}_{\text{r}}=-\frac{6\gamma}{\eta^2 a^2} \; .
\end{equation}
Here $\rho_0$ is the energy-density corresponding to pure dust, i.e., $\rho_0 \propto 1/a^3$. 
The novel contribution $\tilde{\rho}_{\text{r}}$ mimics radiation. The 
solution~\eqref{dust-rad} follows from Eq.~\eqref{consback}, 
where we set ${\cal H} = 1/\eta$. The latter is the standard expression for 
the Hubble parameter during the RD stage. The term $\tilde{\rho}_{\text{r}}$ is somewhat worrisome, 
as it has a negative sign. It governs the evolution of the DM at redshifts 
as large as\footnote{At these early times the notion  of``Modified Dust'' appears 
to be misleading, as the $\Sigma \varphi$-fluid also mimics radiation in that case. However, we choose 
to continue with our standard convention in what follows.} $z \gg z_0 \sim 10^{12}-10^{13}$. 
The value $z_0$ corresponds to the point, when $\dm (z_0)=0$. The temperature of the Universe 
at these early times reads $T \sim 100~\mbox{MeV}-1~\mbox{GeV}$. As these 
temperatures have taken place in the hot Big Bang cosmology, we conclude that 
$\dm <0$ deep in the RD epoch. This, however, 
does not imply the violation of the weak energy condition. 
As pointed out in the previous 
Section, $2\Sigma$ is not the physical energy-density. The latter is given 
by $\dm^{\text{ph}} =T^{0}_{0}$ and remains positive. 
Moreover, due to the time-dependence of the term $\tilde{\rho}_{\text{r}}$, it can 
be absorbed into the standard radiation with no physical consequences
at the background level. 

Let us discuss the linear perturbations around the 
background~\eqref{dust-rad}. At very early times 
corresponding to the redshift values $z \gg z_0$ all the relevant modes are 
beyond the horizon. In the super-horizon regime, i.e., when $k \rightarrow 0$, 
the conservation equation for the DM~\eqref{eq:conservation_equation_term1} takes the form, 
\begin{equation}
\nonumber 
(a^3\delta \dm)'+6\gamma a\left({\cal H}''-{\cal H}{\cal H}'-{\cal H}^3 \right)\Phi_i=0 \; .
\end{equation}
The solution to this equation reads simply
\begin{equation}
\label{supersol}
a^3\delta \dm=C'+\frac{12\gamma a(\eta)}{\eta^2} \Phi_i \; ,
\end{equation}
where $C'$ is the constant in time, which will be specified below. 
At very early times, formally as $\eta \rightarrow 0$, 
the second term on the r.h.s. is the most relevant. In the same limit 
$\dm \rightarrow -6\gamma /\eta^2 a^2$. Consequently, 
we obtain for the energy-density contrast 
at very early times $\delta_{\text{DM}}=-2\Phi_i$,---the standard 
adiabatic initial condition for radiation. As a cross check of our calculations, 
we observe that this initial condition is consistent 
with the $\eta \rightarrow 0$ limit of the $(00)$-component of Einstein's 
equations, see Eq.~\eqref{eq:00_component_term1} of Appendix~\ref{sec:term-gamma_1}.

To conclude, at very early times Modified Dust cannot be distinguished from radiation. 
This is true at both levels of the background and linear perturbations. The things are different for 
sufficiently late times, i.e., at the redshifts $z \ll z_0$. In that case, 
Modified Dust takes the standard form, i.e., $\dm \propto 1/a^3$. 
The first term on the r.h.s. of Eq.~\eqref{supersol} 
corresponds to the constant mode of the energy-density contrast 
$\delta_{\text{DM}}$, while the second 
term stands for the decaying mode. Neglecting the latter, we get the initial condition 
for Modified Dust, $\delta_{\text{DM}} =C'/\rho_{\text{DM},0}$, where
$\rho_{\text{DM},0}$ is the present energy density of DM.  
We fix the constant $C'$ in such a way that we do not encounter 
problems with CMB observations. That is, we 
impose $\delta_{\text{DM}}=-3/2 \Phi_i$ initially. 

A comment is in order before we proceed. Working in terms of the quantity 
$\delta_{\text{DM}}$ can be misleading at the redshift $z_{0}$, 
when $\dm (z_0)=0$. At that point, we 
have $\delta_{\text{DM}} (z_0) \rightarrow \infty$. This, we believe, 
is just a formality, since the quantity $\delta \dm$---more 
relevant one--remains 
finite at all times. To avoid the problem, we could always provide the 
calculations in terms of $\delta \dm$ and then convert the 
latter into $\delta_{\text{DM}}$. 

\subsection{Sub-horizon evolution}

During the RD stage, cosmological perturbations follow a non-trivial evolution, 
which is particularly prominent for short wavelength modes. At sufficiently early times, 
the main contribution to the gravitational potential $\Phi$ follows from the 
perturbations of radiation. In that case, the potential $\Phi$ is given by~\cite{2008cmb..book.....D}
\begin{equation}
\label{potentialrad}
\Phi =-\frac{3\Phi_i}{(u_s k\eta)^2} \left(\cos (u_sk\eta) 
-\frac{\sin (u_sk\eta)}{u_sk \eta} \right) \; ,
\end{equation}
where $u_s=1/\sqrt{3}$ is the sound speed of radiation.
The evolution of the energy-density perturbations can then be 
obtained in a straightforward manner by integrating 
the equation~\eqref{eq:conservation_equation_term1}. We write down the solution,
\begin{eqnarray}
\label{conservationsol} 
\delta_{\text{DM}} =C -9\Phi_i \ln (u_sk\eta) 
+\frac{9\gamma k^2 a(\eta)}{\rho_{\text{DM},0}} \Phi_i +\mbox{decaying modes} .
\end{eqnarray}
The first two terms on the r.h.s. represent the well-known constant and 
logarithmically growing mode in CDM, while the third one is the novelty. 
It describes a mode linearly growing with the scale factor. 
This originates from the term $\sim k^4$ in the conservation 
equation~\eqref{eq:conservation_equation_term1}. 
\begin{figure}[t!]
\begin{center}
\includegraphics[width=0.70\columnwidth]{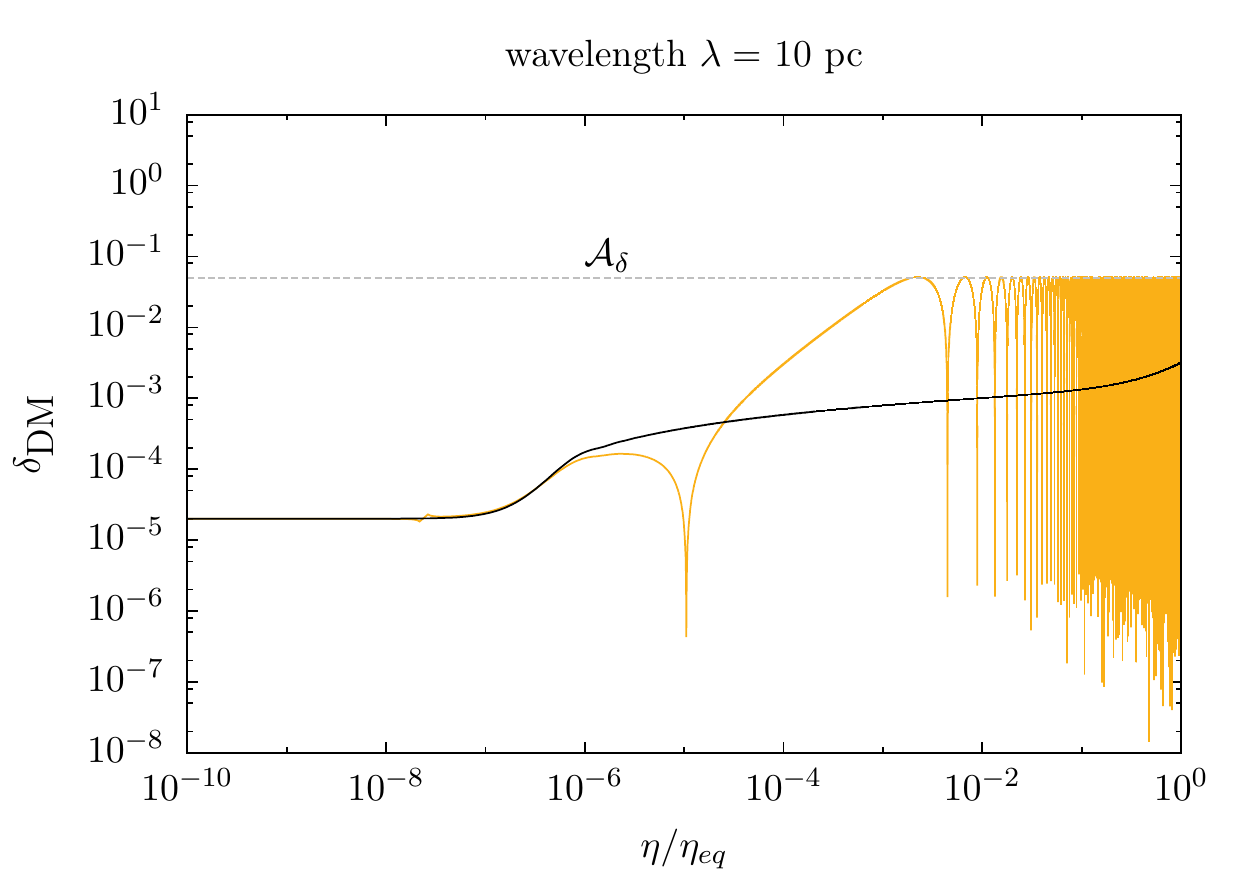}
\end{center}
\caption{The evolution of the energy density contrast for a mode with the wavelength $\lambda =10$ pc 
as a function of conformal time $\eta$. Black and orange lines 
correspond to the cases of a pressureless perfect fluid/CDM 
and Modified Dust, respectively. The choice of the parameter $\gamma=10^{-10} \M^2$ is assumed.
\label{density}}
\end{figure}
While for long wavelength modes this term is negligble, it may start to dominate for shorter ones 
at some point. This point $\eta_{\times}$ in the linear evolution is defined from
\begin{equation}
\label{linearcondition}
\frac{\gamma k^2 a(\eta_{\times})}{\rho_{\text{DM},0}} \sim \ln (u_s k\eta_{\times}) \; .
\end{equation}
Note that the effect takes place only for momenta larger than
\begin{equation}
\label{kmin}
\frac{k}{H_0} \sim  \frac{\M}{\sqrt{\gamma}} \sqrt{z_{eq}} \; ;
\end{equation}
otherwise, the condition~\eqref{linearcondition} formally takes 
place at the MD stage , i.e., at $\eta_{\times} >\eta_{eq}$, 
when the formula~\eqref{potentialrad} and, 
consequently, the estimate~\eqref{kmin} are not applicable. 
Here $z_{eq}$ and $\eta_{eq}$ denote the 
redshift value and the conformal time at the equilibrium between
radiation and matter, respectively, $z_{eq} \sim 10^{4}$ and 
$\eta_{eq} \sim 10~\mbox{Mpc}$. For the value $\sqrt{\gamma}/\M \sim 10^{-5}$, 
we have $k/H_0 \gtrsim 10^{7}$, which corresponds to wavelengths
$\lambda \lesssim 1 ~\mbox{kpc}$. Perturbations with these wavelengths 
are expected to be amplified by the end of the RD stage.

The linear growth continues until the point in the 
evolution, when the DM perturbations become the dominant source of the gravitational potential. 
The associated redshift value $z_*$ is defined from 
\begin{equation}
\label{turn}
\delta_{\text{DM}} \sim \frac{6z_*}{z_{eq}} \Phi_i \; .
\end{equation}
In this estimate, we took into account that perturbations of radiation do not grow, but 
experience oscillations with the amplitude $ 6 \Phi_i$~\cite{Gorbunov:2011zzc}. 
Since this point on, the formula~\eqref{potentialrad} is not applicable anymore. 
To handle the situation, 
let us get back to the conservation equation~\eqref{eq:conservation_equation_term1}, where we 
omit all the terms except for one proportional to $k^4$. Expressing the energy density contrast 
$\delta_{\text{DM}}$  through the perturbations of the velocity potential $\delta \vp$, 
substituting $\delta \vp$ into Eq.~\eqref{eq:00_component_term1} and 
omitting the subleading terms, we obtain the second order equation for the density perturbations, 
\begin{equation}
\nonumber 
\delta''_{\text{DM}}+ 4\pi G \gamma k^2 \delta_{\text{DM}} =0 \; .
\end{equation}
This equation has only 
oscillatory solutions. That is, the growth of perturbations 
$\delta_{\text{DM}}$ stops and we get back 
to the picture observed in the matter dominated Universe. To quantify the amplitude of the 
oscillations, we first estimate the redshift value $z_*$. By comparing Eq.~\eqref{conservationsol} with the 
estimate~\eqref{turn}, we obtain 
\begin{equation}
\nonumber 
z_* \sim  5 \sqrt{z_{eq}}\cdot \frac{\sqrt{\gamma} }{\M} \cdot \frac{k}{H_0}\; .
\end{equation}
Substituting this back into Eq.~\eqref{turn}, we have for 
the amplitude of the energy-density perturbations,
\begin{equation}
\nonumber 
\mathcal{A}_{\delta} \sim \frac{30}{\sqrt{z}_{eq}} \cdot \frac{\sqrt{\gamma}}{\M} 
\cdot \frac{k}{H_0}  \Phi_i \; .
\end{equation}
In Fig.~\ref{density}, we plot the result of numerical simulations for the evolution 
of the energy density contrast $\delta_{\text{DM}}$. As it is clearly seen, it agrees well with our estimates. 
The associated gravitational potential also gets enhanced compared to the 
predictions of CDM scenarios, as it is sourced by perturbations of Modified Dust 
already at redshifts $z \gg z_{eq}$. 
This enhancement gets compensated for the wavelengths $\lambda \gtrsim 1~\mbox{pc}$, since the gravitational potential 
remains constant in the CDM picture at the MD epoch, while it drops with the scale factor 
in the Modified Dust scenario. 

Interestingly, for very short wavelengths, $\lambda \lesssim 1~\mbox{pc}$, the 
energy density contrast reaches unity already during the RD stage.  
We are led to the picture where the Universe becomes inhomogeneous 
at tiny scales already at large redshifts. This picture is in contrast to more common 
WDM and even CDM scenarios, where very small scale perturbations 
are washed out by free-streaming processes. In the future, it would be interesting to see, if that 
phenomenon implies any relevant consequences for observations. 
Though the pc-range of wavelengths is far out of reach for  cosmological 
experiments, it may have important applications for the formation of primordial 
black holes, gravitational lensing, etc. This discussion is out of the scope of the present paper.

Before we finish the discussion about the evolution in the RD background, let us make a comment.  
We remind that generically perturbations are washed out below 
the free-streaming wavelengths of the photon. Naively, 
this effect can be a threat to the mechanism we discussed in this Section. Let us 
show that this never happens in fact. 
Indeed, the comoving free-streaming wavelength $\lambda_{fs}$ drops as $1/z^2$ with the redshift $z$. 
We are interested in the value of $\lambda_{fs}$ at the times, when DM starts to dominate the 
linear evolution. This can be inferred from the value of $\lambda_{fs}$ at the recombination epoch, 
\begin{equation}
\nonumber 
\lambda_{fs} (z_*) \sim \frac{z^2_{rec}}{z^2_*} \lambda_{fs} \sim 10^{9} 
\lambda^2 H^2_0 \cdot 1~\mbox{Mpc} \; .
\end{equation}
Here $z_{rec} \simeq 1000$ is the redshift corresponding to the recombination epoch. 
As it is clearly seen, for the perturbations with the wavelengths 
$\lambda \ll 10^{-5} H^{-1}_0 \sim 100~\mbox{kpc}$, $\lambda_{fs} (z_*) \ll \lambda$. 
Hence, one can neglect the free-streaming effects for all interesting wavelengths.

\section{Non-linear level}
\label{sec:non-linear-level}
Finally, let us discuss the behaviour of Modified Dust in the non-linear regime, i.e., when $\delta_{\text{DM}} \gtrsim 1$. 
Generically, non-linear analysis is a very complicated task, involving non-trivial 
numerical simulations. We leave this for the future work. 
What one can actually do at the moment is to write the relevant system of equations 
in the Newtonian limit, and discuss their possible physical consequences.

Before that, let us remind the state of affairs with the 
collisionless particle DM. In that case, one normally runs the N-body 
simulations, which attempt to solve the system of Vlasov--Poisson equations. 
The problem can be paraphrased in terms of the 
momenta of the Vlasov equation (collisionless Boltzmann equation). 
Formally, it leads to an infinite chain of coupled equations. 
Setting to zero all the momenta of the probability density starting from the velocity dispersion, 
results into the set of equations of the standard dust. 
That is, the conservation equation, 
\begin{equation}
\nonumber 
\dot{\rho} +3H\rho +\frac{1}{a} \nabla \cdot (\rho {\bf v}) =0 \; .
\end{equation}
and the Euler equation 
\begin{equation}
\label{Eulerstandard}
\dot{{\bf v}} +H{\bf v} +\frac{1}{a} ({\bf v}\cdot \nabla) {\bf v} =-\frac{1}{a} \nabla \Phi \; .
\end{equation}
In the present Section, we omit the subscript ``DM'' in the notation of 
the energy density of Modified Dust. This system coadded with the Poisson equation 
can be solved iteratively using Eulerian or Lagrangian perturbation schemes~\cite{Bernardeau:2001qr, Bernardeau:2013oda}. 
In the mildly non-linear regime the results are argued to be in a good 
agreement with N-body simulations. Furthermore, peaks of the energy density localized 
at the surface of particle crossing (caustics, or Zel'dovich pancake) 
give a qualitatively correct picture of the Cosmic Web. Since this point on, 
however, the dust approximation 
breaks down. That is, the divergence of the 
velocity and the energy density become infinite at the caustics, 
i.e., $\nabla \cdot {\bf v} \rightarrow -\infty$ and $\rho \rightarrow \infty$
\footnote{Strictly speaking, this type of divergence is seen in the 
Zel'dovich approximation. It is argued, however, 
that the latter becomes exact for particular initial conditions. Moreover, 
corrections to the Zel'dovich approximation do not cure the problem.}. 
The singularity has a clear physical meaning: it reflects the fact that the 
dust particles may pass unaffected through each other. After particles cross, 
they fly away from the caustics, as there is no mechanism sticking them together. 
This leads to fast broadening of the Zel'dovich pancake, and eventually to diluting the 
structures~\cite{Sahni:1995rm}. The problem 
does not occur in the case of the real particle DM: near the caustics high momenta of the 
Vlasov equation become relevant, and they regularize the divergence. Qualitatively, this amounts 
to the appeareance of an effective pressure (revealed in the non-zero velocity dispersion), 
which opposes gravity, preventing particles to cross. 
Below we argue that a similar mechanism can be relevant in the case of Modified Dust.

We restrict the discussion to the Universe filled in with Modified Dust. 
We write the system of cosmological 
equations in the Newtonian limit. The analogue of the Poisson equation takes the form, 
\begin{equation}
\label{potmod}
\Delta \Phi=\frac{4\pi G}{1-16 \pi \gamma G} a^2\left[\delta \rho- \frac{2H}{a^2}\gamma \Delta \delta \varphi -\frac{\gamma}{a^2} (\nabla \cdot {\bf v})^2 \right] \; .
\end{equation}
The conservation equation reads now
\begin{equation} 
\label{conservnontriv}
\dot{\rho}+3H\rho+\frac{1}{a}\nabla\cdot (\rho {\bf v}) 
= -\frac{\gamma}{a^3} \Delta (\nabla \cdot {\bf v})  \; .
\end{equation}
The Euler equation has the standard form~\eqref{Eulerstandard}, 
as it follows from the constraint equation~\eqref{constraint}. Here ${\bf v}$ 
is the velocity defined with respect to the Euclidean space. 
It is related to the velocity potential $\varphi$ by ${\bf v} =-\nabla \varphi /a$.

One can show that the modification of the Poisson equation is irrelevant. 
Indeed, the extra terms in Eq.~\eqref{potmod} result into small 
${\cal O} \left( \frac{\gamma}{M^2_{Pl}} \right)$ corrections to the 
terms already present in the Euler equation~\eqref{Eulerstandard}. 
At the same time, the modification of the conservation equation is something new compared 
to the standard dust. Interestingly, deviations from the latter 
start explicitly when the different trajectories come very close to each other, 
so that the quantity $\nabla \cdot {\bf v}$ tends to blow up, 
i.e., $\nabla \cdot {\bf v} \rightarrow -\infty$. 
In that case, the r.h.s. of 
Eq.~\eqref{conservnontriv} becomes sufficiently large. 
This gives rise to the flow of energy away from the region, where one would 
expect the presence of singularities, to the outer regions. As nothing prevents the energy 
density from becoming negative, one results with  ``anti-gravity'', 
i.e., a repulsive force between fluid elements. This is a necessary---though, 
not sufficient,---condition to cure the caustic singularities. Note that the appearence of the negative 
energy density does not imply catastrophic ghost instabilities in the model. 
Indeed, as it follows from Eq.~\eqref{conservnontriv}, the 
integral $\int \rho dV$ is conserved for any large physical volume $V$. 
Hence, the regions of negative energy do not swallow the entire space. 

To illustrate the picture described above, let us consider the toy example 
of the 1D collapse. Namely, we take some initial sufficiently smooth distribution of the 
energy density $\rho (t=0,x)=A \mbox{exp} \left( -\frac{x^2}{2L^2} \right)$ with the initial 
velocity $v(t=0,x)=0$. Here $A$ is the constant amplitude and $L$ is the characteristic size of the distribution. Let us
focus on the case $\gamma=0$. We assume in what follows 
that the Universe is empty, and the scale factor $a=1$. The solution for the 
energy density reads in the approximation $|x| \ll L$,
\begin{equation}
\label{rhosing}
\rho (t,x)=\frac{A}{1-2\pi G A t^2}-\frac{Ax^2}{2(1-2\pi GA t^2)^4 L^2} +{\cal O} (x^4)\; .
\end{equation}
In the same approximation, the solution for the velocity is given by
\begin{equation}
\label{velsing}
v (t,x)=-\frac{4\pi G Axt}{1-2\pi GAt^2}+\frac{2\pi G Ax^3t}{3(1-2\pi GAt^2)^4 L^2} +{\cal O} (x^5)\; .
\end{equation}
While we are primarily interested in the behaviour at $x=0$, where the appearance of the singularity is expected, 
we write explicitly ${\cal O}(x^2)$ corrections for 
future purposes. As it is clearly seen, the energy density blows up at a finite time $t_s=1/\sqrt{2\pi GA}$. 
The same happens to the divergence of the velocity, i.e., $\partial_x v \rightarrow -\infty$. This is the caustic singularity. 
Now let us include the $\gamma$-term into the discussion. 
Still assuming that the solution~\eqref{rhosing} and~\eqref{velsing} holds, we obtain for the r.h.s. of 
Eq.~\eqref{conservnontriv} at $x=0$, 
\begin{equation}
\label{higher}
-\gamma \partial^3_x v=-\frac{4\pi \gamma G A t}{(1-2\pi GAt^2)^4 L^2} \; .
\end{equation}
As it follows, the higher derivative term becomes of the order of the standard dust term in Eq.~\eqref{conservnontriv} 
even for an arbitrarily small $\gamma$ and large $L$ at times sufficiently close to $t_s$. Another important fact is that
the $\gamma$-term has a negative sign. Therefore the growth of the energy density slows down. 
Since this point on, however, the solutions~\eqref{rhosing},~\eqref{velsing} and, 
consequently,~\eqref{higher} are not valid anymore, as they were obtained in the assumption of a negligible $\gamma$-term. Assuming, however, that 
the higher derivative term becomes dominant, one observes the 
decrease of the energy density. As the positive values of the energy 
density are not protected, $\rho$ may become negative at some point. 
This is in agreement with the qualitative picture described above. 
One may worry that the energy density will continue to decrease infinitely. 
This, we expect, does not happen, since the case $\rho < 0$  corresponds to a 
repulsive force between fluid elements accordingly to Eqs.~\eqref{potmod} and~\eqref{Eulerstandard}. 
Therefore, the velocity and presumably higher derivative term on the 
r.h.s. of Eq.~\eqref{conservnontriv} change their sign at some point. 
As a result, the decrease of the energy density stops and turns into growth. 
If so, there is a natural mechanism for stabilizing the Cosmic Web.

At the moment, we are carrying out numerical simulations of the gravitational collapse of Modified Dust~\cite{caustic2015}. 
Preliminary results confirm the picture described above for the range of 
parameters $\gamma /A L^2 \gtrsim 1$. In a more interesting case $\gamma /A L^2 \ll 1$, however, 
the solution is the subject of instabilities, which can be either due to some 
numerical artefacts or the presence of caustic singularities.  
A study of this part of the parameter space is currently under way.

Given that this mechanism indeed works, one deals 
with a novel way of curing caustic singularities. 
Recall that the common lore is to modify the Euler equation. Namely, 
one adds viscous terms to the r.h.s. of Eq.~\eqref{Eulerstandard} as 
in adhesive gravitational models~\cite{Gurbatov:1989az, Buchert:2005xj}, 
or ``quantum'' pressure models~\cite{Widrow:1993qq} (see also~\cite{Uhlemann:2014npa} 
for the recent studies). These attempts amount to parametrizing the velocity dispersion 
present in the particle DM case. 
Therefore, we may expect that the Modified Dust scenario leads to qualitatively new results.

In particular, the higher derivative term could be 
promising for solving the other long-standing problem of the CDM: the cuspy profile of the 
DM halos. To show this, let us estimate the distance from the centre of the 
dwarf spheroidal galaxies, 
at which the source term on the r.h.s. of Eq.~\eqref{conservnontriv} becomes 
relevant: 
\begin{equation}
\nonumber
\frac{|\gamma \Delta (\nabla \cdot {\bf v}) |}
{| \nabla (\rho \cdot {\bf v}) |} \sim \frac{\gamma}{\rho r^2} \gtrsim 1 \; ,
\end{equation}
We estimate the energy density of DM in the central regions of the galaxies as 
$\dm \sim 10~\mbox{GeV}/{\mbox{cm}^3}$. 
We use the estimate for the parameter $\gamma \sim 10^{-10} \M^2$, 
as it is the most relevant for the solution of the 
missing satellites problem. Then, the inequality 
above gives 
\begin{equation}
\label{estimate}
r \lesssim 100~\mbox{pc} \; .
\end{equation}
This roughly corresponds to the scales, at which the density profiles obtained in simulations begin to 
disagree with the observational data~\cite{2010AdAst2010E...5D}. 
At distances smaller than about $100$ pc, we expect a substantial flow of energy away 
from the centre. Hence, one has a chance to reduce the mass of 
DM in the central region. 
 
We reiterate that all the conclusions made in this Section are preliminary. 
By no means, they should be viewed as the proof of absence of caustic 
singularities or interpreted as the successful 
solution to the core-cusp problem. 
This remains to be shown by making use of numerical simulations. 
We leave this for a future work.

\section{Conclusions}
In this paper we showed that the Modified Dust scenario 
allows to address several cosmological puzzles. For rather long wavelengths and relatively 
low values of the parameter $\gamma$, cosmological perturbations behave as in the CDM picture. 
Below some wavelength, the power spectrum 
gets suppressed compared to the standard predictions. This could be relevant for alleviating 
the missing satellites problem. In the previous Section, we also showed that 
Modified Dust has some appealing features in the Newtonian limit. That is, 
unlike the standard dust, it may avoid developing singularities at the caustics.

In this regard, the scenario which we considered in the present paper is a good model for DM. 
Before making that strong statement, however, several important issues must be addressed:

\begin{itemize}
\item Numerical simulations must be performed in the non-linear regime. 
In particular, it would be interesting to see if caustic singularities are indeed absent. 
In the case of a
positive answer, one can ask more sophisticated questions. For example, 
what is the structure of DM halos formed by Modified Dust?

\item Lyman-$\alpha$ forest data is a powerful tool for descriminating between different DM 
frameworks. Therefore, it is important to test the predictions of the Modified Dust scenario using these data, and 
possibly to deduce the constraints on the parameter $\gamma$. 
 
\item Generically, suppression of sub-galactic scale structures implies the delayed formation 
of the first stars compared to CDM predictions. This may lead to some tension with the CMB data that favors an early 
reionization of the Universe. The situation looked hopeless with the first release of 
the WMAP data~\cite{Yoshida:2003rm,Somerville:2003sh}, which reported 
large redshifts values $z_{re}$ corresponding to the 
half-reionized Universe~\cite{Kogut:2003et}: $11<z_{re} < 30$ at 95\% C.L.. 
However, the best-fit value of the redshift $z_{re}$ essentially decreased with the later 
releases of WMAP data and Planck data~\cite{2013ApJS..208...19H,2013arXiv1303.5076P}. Hence, one has 
a chance to avoid stringent constraints on the parameter $\gamma$. 

\item Is the model under study consistent with the Bullet Cluster observations? 
This issue concerns the non-linear dynamics of Modified Dust and 
thus remains obscure at the moment.

From the theoretical point of view, there are the following issues: 

\item The unified description of the dark matter and dark energy. 
Recall that this has been the original motivation of the 
$\Sigma \varphi$-fluid. One can try to address this issue following the guidelines of the 
paper~\cite{Lim:2010yk}. 

\item It would be certainly worth to search for a fundamental theory underlying 
Modified Dust. In particular, higher derivative terms are naturally viewed as the 
part of an effective theory associated with some broken global symmetry, with 
$\varphi$ being the Goldstone field. On the other hand, this analogy with the effective theory 
is complicated by the presence of the constraint~\eqref{constraint}. 

Once a solution to these problems is found, it would be interesting to search 
for signatures of the Modified Dust scenario in the observational data.

\end{itemize}
\textit{Note added.} At the final stage of this project, we became aware of the related work carried out 
by L.~Mirzagholi and A.~Vikman. The discussion of Ref.~\cite{Mirzagholi:2014ifa} made available recently essentially extends 
that of the present paper.

\subsubsection*{Acknowledgments}

We thank Dmitry Gorbunov, Michael Gustafsson, Mikhail Ivanov, Andrey Khmelnitsky, Maxim Pshirkov,
Tiziana Scarna, Sergey Sibiryakov, Peter Tinyakov and Alexander Vikman for many useful 
ÃÂcomments and fruitful discussions. We are indebted to 
A.~Vikman for sharing several ideas of the work~\cite{Mirzagholi:2014ifa} prior to its publication. This work is supported by the 
Wiener Anspach Foundation (F.~C.) and Belgian Science Policy IAP VII/37 (S.~R.).

\appendix
\section{Term $\frac{\gamma_1}{2} (\square \vp)^2$}
In this Appendix, we write the system of 
cosmological equations for the Universe filled in with 
Modified Dust and radiation. As in the bulk of the paper, 
we assume the parameter $\gamma_2=0$.
\label{sec:term-gamma_1}
\subsection{Background level}

We start with background cosmological equations. In the presence of radiation and 
Modified Dust, the Friedmann equation takes the form
\be
3\lb(1-20\pi \G \gamma_1)\mathcal{H}^2 +8\pi \G \gamma_1 \mathcal{H}'\rb= 
8\pi \G a^2\lp \dm+\rad\rp .
\label{eq:Friedmann_appendix}
\ee
Here $\rad$ is the energy density of radiation; we assume the simple identification $2\Sigma=\dm$ as in the 
bulk of the paper. 
The $(ij)$-component of the Einstein's equation is given by
\be
-3\lp1-12\pi \G \gamma_1 \rp \lp  2\mathcal{H}'+\mathcal{H}^2\rp = 8\pi \G a^2 \rad. 
\label{eq:radiation_appendix}
\ee
We supplement the system with conservation equations for radiation
\be
\rho'_r +4{\cal H} \rho_{r} =0
\ee
and Modified Dust
\be
a^2 \dm '+ 3\mathcal{H}a^2 \dm+ 3 \gamma_1 \lp \mathcal{H}^3+
\mathcal{H}\mathcal{H}'-\mathcal{H}'' \rp = 0.
\label{eq:conservation_appendix}
\ee
Taking present values for the energy density of 
radiation and DM, one can easily solve this system. In particular, 
neglecting the contribution for radiation, 
we get back to the system of equations 
given in Section~\ref{sec:einstein-de-sitter}.

\subsection{Linear level}
The simplest equation at the linear level follows from the constraint 
$g^{\mu\nu}\partial_{\mu}\vp\partial_{\nu}\vp=1$, which, we remind, 
remains unmodified upon the inclusion of $\gamma$-terms. Namely, 
it is given by Eq.~\eqref{euler}. We repeat it here for the sake of completeness,
\be
\delta \vp '= a \Phi. 
\label{eq:constraint_equation_appendix}
\ee
The $(00)$-component of the Einstein's equations reads in Fourier space,
\bea
\nonumber
&-k^2 \Phi \left(1 -4\pi \gamma_1 \G \right) -3\Phi 
\left({\cal H}^2 +4\pi \gamma_1 \G \left[ 2{\cal H}'-5{\cal H}^2\right]\right)-
\\
&3{\cal H} \Phi' (1-12\pi \gamma_1 \G)-12 \pi \gamma_1 \G \Phi''
=4\pi \G \left( a^2 \rho_{\text{r}} \delta_{\text{r}}+a^2 \dm \delta_{\text{DM}} 
+\frac{5 {\cal H} \gamma_1}{a} k^2 \delta \varphi 
\right)\; ,
\label{eq:00_component_term1}
\eea
where $\delta_{\text{r}}$ is the perturbation for the radiation energy density.
The $(0i)$-component is given by: 
\bea
 -3\lp 1-12 \pi \G \gamma_1 \rp \delta \vp''
+12\pi \G \lb 3\gamma_1\lp\mathcal{H}^2-\mathcal{H}'\rp+a^2\dm
-\gamma_1 k^2 \rb \delta \vp &=& -16\pi \G a^2\rad \vrad ,
\eea
where $\vrad$ is the scalar velocity potential for radiation; 
here we exploited Eq.~\eqref{eq:constraint_equation_appendix} to 
express the gravitational potential $\Phi$ through $\delta \varphi$.  
Neglecting the energy density of the radiation in the last 
equation, we get back to Eq.~\eqref{0iapprox}
from the main body of the paper with the sound speed $c^2_s$ given by
\be
c^2_s=\frac{4\pi G \gamma_1}{1-12 \pi \gamma_1 G}.
\label{eq:close_equation_vp_term1}
\ee
The $(ij)$-component of Einstein equations is given by
\bea
3(1-12\pi \G \gamma_1)\left\{(\mathcal{H}^2-\mathcal{H}') \delta \vp'
-\mathcal{H}\delta \vp''-\delta \vp''' \right\}
-12\pi \G \gamma_1 k^2 \lp \delta \vp'+ \mathcal{H}\vp \rp  = -4\pi \G a^3 \delta \rad. \nonumber 
\eea
Note that the non-diagonal part of the same equation equals to zero identically. This, 
we remind, is a property of the $\gamma_1$-term, which preserves the simple form of the energy-momentum tensor 
$\delta T^i_j\propto \delta^i_j$. This similarity with the case of the perfect dust breaks down at the 
level of the $\gamma_2$-term, as we explain in details in the following Appendix. 

The system supplemented with the conservation equation for Modified Dust
\be
\lp a^3 \delta \dm \rp'-3\gamma_1 \lp\mathcal{H}''-\mathcal{H}\mathcal{H}'
-\mathcal{H}^3 \rp \delta \vp'+\dm k^2 a^2 \delta \vp= 3\Phi' a^3 \dm+
a^4 \gamma_1 \left.\lp\square^2 \vp \rp \right|_{\text{linear}}
\label{eq:conservation_equation_term1}
\ee
where
\bea
\left.\lp\square^2 \vp \rp \right|_{\text{linear}}=
 & =\frac{1}{a^4} [(k^4-2{\cal H}'k^2)\delta \varphi 
+(18{\cal H} {\cal H}'-6{\cal H}''-6{\cal H}^3-2{\cal H}k^2) \delta \varphi'
\nonumber
\\ & + (12{\cal H}^2-6{\cal H}'-2k^2)\delta \varphi''+
6{\cal H} \delta \varphi'''-3 \delta \varphi''''] \; 
\eea
and the standard conservation equation for the radiation 
can be solved numerically with the initial conditions set 
deep in RD stage (see discussion in Section~3).

\section{Term $\frac{\gamma_2}{2} \nabla_{\mu}\nabla_{\nu}\vp \nabla^{\mu}\nabla^{\nu}\vp$}
\label{sec:term2}
Now, let us consider the effects of the $\gamma_2$ term on the 
cosmological evolution. That is, we set the parameter $\gamma_1$ to zero. 
We restrict the discussion to the 
case of the matter dominated Universe filled in 
with Modified Dust. Namely, we neglect the contribution 
of the radiation. 

\subsection{Background level}

First, let us show that the background cosmological equations are the same 
as in the case of a pressureless perfect fluid. This immediately 
follows from the $(ij)$ component of the 
Einstein equation, which has the standard form $2{\cal H}'+{\cal H}^2=0$. 
The Friedmann equation is given by
\be
3 \lp1-12\pi \G \gamma_2 \rp \mathcal{H}^2 = 8\pi \G a^2 \dm
\label{eq:Friedmann_term2}
\ee
and the conservation equation reads
\be
a^2 \dm' + 3a^2 \mathcal{H}\dm + 3\gamma_2 \mathcal{H} 
\lp 2\mathcal{H}'+\mathcal{H}^2 \rp = 0.
\label{eq:conservation_equation_term2}
\ee
We see that the quantity $\dm \equiv 2\Sigma$ drops as $1/a^3$ with the 
scale factor. Hence, it can be identified with the 
energy density of DM, as in the case of 
the $\gamma_1$-term. 

\subsection{Linear level}
Let us discuss the non-trivial effects for the linear cosmological 
perturbations due to the presence of the $\gamma_2$-term. 
In that case, the simple relation $\delta T^{i}_{j} \propto \delta^{i}_{j}$ 
does not hold anymore. Therefore, the gravitational potentials 
$\Phi$ and $\Psi$ do not coincide. Still, the main conclusion of the paper holds: cosmological 
perturbations with sufficiently short wavelengths are suppressed.

To show this explicitly, we write down the 
$(00)$ and $(ij)$-components of the Einstein equations,
\be
4\pi \G\lp a^2 \dm +3\gamma_2 \mathcal{H}^2 - \gamma_2 k^2 \rp \delta \vp -\mathcal{H} 
\lp 1-4\pi \G  \gamma_2 \rp \delta \vp' 
-a \lp1-4\pi \G \gamma_2 \rp \Psi'= 0,
\ee
and
\bea
&\delta_i^j \left[ ak^2 \Psi +2a(1-4\pi\G\gamma_2)(2\mathcal{H}\Psi'+\Psi'')
+4\mathcal{H}'(1-4\pi\G \gamma_2)\delta \vp'-k^2 \delta 
\vp'+\right.\nonumber \\
&+\left. 2\mathcal{H}(1-4\pi \G \gamma_2)\delta \vp'' \right] 
+k_i k^j \lb 8\pi \G \gamma_2 \mathcal{H} \delta \vp+ (1+8\pi \G \gamma_2) \delta \vp' 
- a \Psi \rb = 0 \; ,
\label{eq:ij_component_term2}
\eea
respectively.  The tensor structure of the 
last equation contains two parts: one proportional 
to $\delta^{i}_{j}$ and the other proportional to $k_i k^j$. The latter 
gives the relation between the potentials $\Phi$ and $\Psi$ ($i\neq j$ case),
\be
a\Psi=8\pi \G \gamma_2 \mathcal{H} \delta \vp+ (1+8\pi \G \gamma_2) \delta \vp'. 
\ee
Using this, we can express the derivatives of the potential $\Psi$ in terms 
of the derivatives of the gravitational potential $\Phi$ and $\delta \varphi$. Plugging 
the previous relation 
into the $(00)$-component of the Einstein's equations, we finally obtain 
\be
\delta \vp'' + \lp \frac{8\pi \G\gamma_2}{2+8\pi\G\gamma_2-(8\pi\G\gamma_2)^2}k^2
-\frac{3}{2}\mathcal{H}^2 \rp \delta \vp = 0
\ee
Here, we made use of the background equation~(\ref{eq:Friedmann_term2}). 
We see that in the limit $\gamma_2 \ll \M^2$, it reduces to Eq.~(\ref{0iapprox}) 
studied in the main body of the paper. 
We conclude that all 
the results studied for the case of the $\gamma_1$-term in Section~\ref{sec:einstein-de-sitter}
are true for the case of the $\gamma_2$-term.

\bibliographystyle{apsrev}
\bibliography{refs}

\end{document}